\documentclass[a4paper]{article}
\usepackage[dvipsnames]{xcolor}
\definecolor{mygray}{gray}{0.6}
\usepackage{INTERSPEECH2021}
\usepackage{cite}
\usepackage{bbding}
\usepackage{hyperref}

\title{NeuralEcho: A Self-Attentive Recurrent Neural Network For Unified Acoustic Echo Suppression And Speech Enhancement}
\name{Meng Yu, Yong Xu, Chunlei Zhang, Shi-Xiong Zhang, Dong Yu}
\address{
  Tencent AI Lab, Bellevue, WA, USA}
\email{\{raymondmyu, lucayongxu, cleizhang, auszhang, dyu\}@tencent.com}

\begin{document}

\maketitle
\begin{abstract}
Acoustic echo cancellation (AEC) plays an important role in the full-duplex speech communication as well as the front-end speech enhancement for recognition in the conditions when the loudspeaker plays back. In this paper, we present an all-deep-learning framework that implicitly estimates the second order statistics of echo/noise and target speech, and jointly solves echo and noise suppression through an attention based recurrent neural network. The proposed model outperforms the state-of-the-art joint echo cancellation and speech enhancement method F-T-LSTM in terms of objective speech quality metrics, speech recognition accuracy and model complexity. We show that this model can work with speaker embedding for better target speech enhancement and furthermore develop a branch for automatic gain control (AGC) task to form an all-in-one front-end speech enhancement system.

\end{abstract}

\section{Introduction}
The acoustic echo is caused by the sound from the loudspeaker which is received by the near-end microphone and then transmitted to the far-end listener or speech recognition engine. Such interfering signal degrades the voice quality in teleconference system, mobile communication and hand-free human-machine interaction \cite{Benesty01, Enzner14}. In addition, the environmental noises also degrade the audio quality and as well as limit the performance of AEC algorithms \cite{Gustafsson02}. 
Adaptive filtering methods have been studied over decades for estimating the acoustic echo path and linear echo cancellation \cite{Soo90, Valin07}. Normalized least mean square (NLMS) is most widely used due to its robustness and low complexity, such as frequency domain block adaptive filter (FDBAF) and the multi-delay block frequency domain (MDF) adaptive filter \cite{Soo90}. The non-linear post processing is usually cascaded for residual echo suppression (RES). However, these methods are not efficient for non-linear echo distortion, echo path change and non-stationary noises \cite{Turbin97, Hansler05}. Inspired by the great success of deep learning, the deep neural networks have been developed to address non-linear echo suppression problem. Networks such as complex-valued DNNs \cite{Halimeh21,Peng21}, Long Short Term Memory networks (LSTM), and multi-head self-attention \cite{Ma20, Ma21}
have been employed to develop echo suppression system for better handling non-linear echo distortions and echo path delay. 
Therefore linear adaptive filtering followed by the neural network based RES is adopted to form a hybrid system for AEC system design \cite{Valin21, Wang21, Ma20}. 
In the recent AEC Challenge \cite{Sridhar21,Cutler21}, the hybrid systems \cite{Valin21}, \cite{Wang21} and \cite{Peng21} achieved promising results. AEC has been formulated as an end-to-end supervised speech enhancement problem in \cite{Zhang18}, where a bidirectional long-short term memory (BLSTM) network was adopted to predict the mask for extracting the near-end target speaker. Such end-to-end AEC model attracts research attention and has been widely developed in recent works. The
dual signal transformation LSTM network (DTLN) was adapted to the AEC task in \cite{Westhausen21}. A Wave-U-Net based acoustic echo cancellation with an attention mechanism was proposed in \cite{Stoller20} to jointly suppress acoustic echo and background noise. The attempt on using complex neural network together with frequency-time-LSTMs (F-T-LSTM) \cite{Zhang21} leads to important phase information modeling and temporal modeling, outperforming the best method in AEC Challenge \cite{Sridhar21}.

Inspired by the two-stage multi-channel joint AEC and beamforming in \cite{Kothapally21}, we propose an end-to-end two-stage single-channel AEC and noise reduction model in this paper, namely NeuralEcho. Specifically, the model implicitly estimates the echo signal and AEC output in the first stage. The echo/noise covariance matrix and target speech covariance matrix are estimated across the estimated echo channel, AEC output channel and the input microphone channel. The speech enhancement filters are then predicted in the second stage given the second order statistics of echo/noise and target speech. 
This model leverages cross-channel second-order statistics for better echo/noise reduction and target speaker preservation. Together with RNN model structure, a multi-head self-attention \cite{Vaswani17} is employed over time to dynamically emphasize relevant features. Furthermore, similar as the feature fusion used in \cite{Malley21}, we incorporate feature-wise linear modulation (FiLM) \cite{Perez18} for combing the target speaker embedding vector with other acoustic features. It allows the network to better separate the target speech signal from echo and other interference. We also integrate AGC \cite{Chu96} to amplify speech signal to an intelligible sound level, which is in high demand by real-time communications (RTC). Wang et al. \cite{Ziteng21} recently presented a NN3A model which supports audio front-end AEC, noise suppression and AGC. Particularly, a traditional AGC algorithm is placed in a post-processing stage. We show that by training NeuralEcho and AGC task in a unified model, the output of AGC branch not only adjusts
the processed signal to a proper sound level, but also improves the speech recognition accuracy. 

\begin{figure*}[!t]
        \centering
        \vspace{-2.2cm}
        \includegraphics[width=\linewidth,height=130mm]{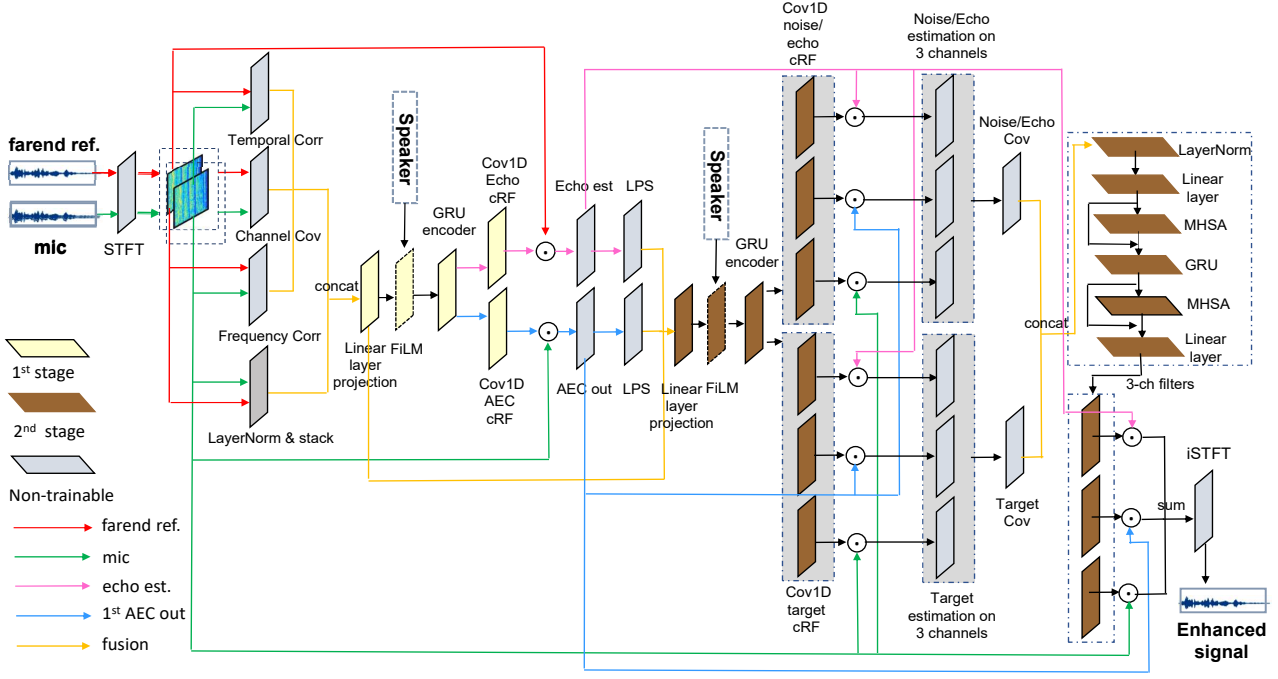}
        \vspace{-2.3cm}
        \caption{The architecture of the proposed NeuralEcho model with speaker information as an option.}
        \label{fig:model}
        \vspace{-0.4cm}
\end{figure*}

The rest of the paper is organized as follows. In Section \ref{sec:neuralecho}, we present the NeuralEcho model, as well as its speaker aware version and the one with AGC module. We describe our experimental setups and evaluate the effectiveness of the proposed system in Section \ref{exp}. We conclude this work in Section \ref{con}.

\section{Proposed Model}\label{sec:neuralecho}
We formulate the problem as enhancing the near-end target speech signal received by a single microphone with the presence of reverberation, far-end echo signal from the loudspeaker, and the background noise. Let $s$ and $x$ represent the dry clean signal from near-end and far-end, respectively. The signal captured by the microphone at time $t$ can be represented by
\begin{equation}\label{signal}
    y(t) = s_r(t) + x_r(t) + v(t),
\end{equation}
where $s_r(t) = h_s(t) * s(t)$ and $x_r(t) = h_x(t) * g(x(t))$ are the reverberant target speech signal and loudspeaker
emitted distorted far-end signal, respectively, $v(t)$ denotes the near-end noise signal and $*$ denotes the convolution. The function $g$ denotes the non-linear distortion, while echo path delay and near-end reverberation are included in $h_x(t)$. This study aims at designing a neural network in Section \ref{sec:neuralecho} to predict the reverberant target speech $s_r(t)$ while suppress the far-end echo signal and background noise given the microphone signal $y(t)$ and far-end reference signal $x(t)$. Furthermore, the target speaker embedding vector $d_s$ serves as a bias condition in Section \ref{sec:spk} and the level adjusted dry clean target speech signal $\mathcal{A}(s(t))$ is estimated through an AGC branch in Section \ref{sec:agc}.

\subsection{NeuralEcho Model}\label{sec:neuralecho}
The audio samples $y(t)$ and $x(t)$ are first transformed to frequency domain as $\mathbf{Y}(k,n)$ and $\mathbf{X}(k,n)$, respectively, by short time Fourier transform (STFT), where $k$ represents frequency bin and $n$ represents frame index. We use 512-point FFT, with 32ms hann window and 16ms hop size. By stacking up $\mathbf{Y}$ and $\mathbf{X}$, we have $\mathbf{Z}(k,n) = [\mathbf{Y}(k,n), \mathbf{X}(k,n)]^T $. The covariance matrix $\mathbf{\Phi}^{C}(k,n)\in \mathbb{C}^{2\times 2}$, denoted by Channel Cov in Fig. \ref{fig:model}, is used as an input feature. This accounts for cross-correlation between the far-end reference signal and the microphone signal,
as well as power spectral density (PSD) of microphone signal and far-end reference signal. $\mathbf{\Phi}^{C}(k,n)$ is computed as 
\begin{equation}\label{cov}
    \mathbf{\Phi}^{C}(k,n) = (\mathbf{Z}(k,n) - \mathbf{\mu}_z)(\mathbf{Z}(k,n) - \mathbf{\mu}_z)^H, \\
\end{equation}
where $\mathbf{\mu}_z  = \frac{1}{2} \sum_{i=1}^2 \mathbf{Z}_i(k,n)$, ‘$i$’ represents the channel index and $(\cdot)^H$ represents Hermitian operator.
We discard the upper half of the complex symmetrical matrix to reduce computational cost and memory usage. The lower triangular matrix is flatten and the real and imaginary parts are concatenated.
Furthermore, for microphone and far-end reference signal, we compute its correlation matrix across time frames and frequency bins separately, namely Temporal Correlation $\mathbf{\Phi}^{T}$ and Frequency Correlation $\mathbf{\Phi}^{F}$ as
\begin{align}
    \mathbf{\Phi}^{T}_i(k,n)  = & [\mathbf{Z}_i(k,n), \mathbf{Z}_i(k,n-1), \dots, \mathbf{Z}_i(k,n-n_{\tau})]  \notag \\ 
    & [\mathbf{Z}_i(k,n), \mathbf{Z}_i(k,n-1), \dots, \mathbf{Z}_i(k,n-n_{\tau})]^H \\
    \mathbf{\Phi}^{F}_i(k,n)  = & [\mathbf{Z}_i(k,n), \mathbf{Z}_i(k-1,n), \dots, \mathbf{Z}_i(k-k_{\tau},n)]  \notag \\ 
    &  [\mathbf{Z}_i(k,n), \mathbf{Z}_i(k-1,n), \dots, \mathbf{Z}_i(k-k_{\tau},n)]^H, \notag
\end{align}
where $i$ represents microphone and far-end reference channels, and $n_{\tau}$ and $k_{\tau}$ correspond to the maximum shift along time and frequency axis, respectively. We set them to 9 in our model for capturing the signal's temporal and frequency dependency. Similarly, the lower triangular matrix is flatten and the real and imaginary parts of both microphone and far-end reference channels are concatenated. The three features $\mathbf{\Phi}^{C}$, $\mathbf{\Phi}^{T}$ and $\mathbf{\Phi}^{F}$ are concatenated, together with the normalized log-power spectra (LPS) of microphone and far-end reference signals. 

In the first stage, the input feature vector for each time-frequency bin is projected to a one-dimensional space through a linear projection layer (from 368-dim to 1-dim). After a GRU encoder layer (257 hidden units), the model estimates
$((2K+1)\times(2N+1))$ dimensional complex-valued ratio filters \cite{Mack19,Zhuohuang21} $\mathbf{cRF}_{AEC}(k,n)$ and $\mathbf{cRF}_{echo}(k,n)$ for the first stage AEC output and echo estimation, respectively. Eq.\ref{crf} demonstrates the computation of applying the estimated $\mathbf{cRF}_{AEC}(k,n)$ on time-frequency shifted version
of input microphone signal $\mathbf{Y}(k,n)$ to produce echo suppressed signal. Similar operation is applied on far-end reference signal $\mathbf{X}(k,n)$ to produce echo estimation $\mathbf{X}_{echo}(k,n)$.
\begin{equation}\label{crf}
        \mathbf{Y}_{AEC} = \sum\mathbf{cRF}_{AEC}(k,n,\tau_k,\tau_n) * \mathbf{Y}(k+\tau_k,n+\tau_n)
        \vspace{-0.3cm}
\end{equation}
\hspace{1cm}${\tau_k\in[-K,K],\tau_n\in[-N,N]}$ \vspace{-0.2cm}\\

In the second stage, we use the raw microphone input channel, first stage AEC processed channel, and echo estimation channel $\Tilde{\mathbf{Z}}(k,n) = [\mathbf{Y}(k,n), \mathbf{Y}_{AEC}(k,n), \mathbf{X}_{echo}(k,n)]^T$ to learn the speech enhancement solution filters from the estimated echo/noise covariance matrix $\mathbf{\Phi}_{NN}(k,n)$ and target near-end speech covariance matrix $\mathbf{\Phi}_{SS}(k,n)$. We first compute the LPS on $\mathbf{Y}_{AEC}$ and $\mathbf{X}_{echo}$, which are then concatenated with the acoustic feature from the linear projection layer in the first stage to serve as the second stage input feature. The input feature goes through a linear projection layer (257 hidden units) and a GRU layer (257 hidden units). The GRU outputs are then passed through one dimensional convolution layers to estimate complex ratio filters ($\mathbf{cRF}_{N}$ and $\mathbf{cRF}_{S}$) for estimating the echo/noise and target speech in $\Tilde{\mathbf{Z}}$. The complex ratio filters are applied to $\Tilde{\mathbf{Z}}$ in the same way as Eq. \ref{crf} to estimate the 3 channels' complex spectrum of echo/noise and target speech, respectively, for each time-frequency bin. To this end, similar to Eq. \ref{cov} we compute the echo/noise covariance matrix $\mathbf{\Phi}_{NN}(k,n)$ and target speech covariance matrix $\mathbf{\Phi}_{SS}(k,n)$, followed by flattening the lower triangular matrix and concatenating the two matrices' real and imaginary parts. These features are normalized through layer normalization \cite{Ba16} and then fed to the self-attentive RNN layers for enhancement filter estimation. The model structure is formulated as 
\begin{align}
    &\mathbf{\Phi}_{norm}  = \mathbf{LayerNorm}([\mathbf{\Phi}_{NN}, \mathbf{\Phi}_{SS}]), \\
    &\mathbf{\Phi}_{proj}  = \mathbf{LeakyReLU}(\mathbf{DNN}(\mathbf{\Phi}_{norm})), \\
    &\mathbf{\Phi}_{attn}  = \mathbf{\Phi}_{proj}(k,n) + \mathbf{ReLU}(\mathbf{MHSA}(\mathbf{\Phi}_{proj})), \\
    &\mathbf{\Phi}_{gru}  = \mathbf{GRU}(\mathbf{\Phi}_{attn}), \\
    &\mathbf{\Phi}_{attn}  = \mathbf{\Phi}_{gru} + \mathbf{ReLU}(\mathbf{MHSA}(\mathbf{\Phi}_{gru})), \\
    &\mathbf{W}  = \mathbf{DNN}(\mathbf{\Phi}_{attn}),
\end{align}
where MHSA stands for multi-head self-attention with 4 parallel attention heads in our setup. We omit the time-frequency index $(k,n)$ in the above formula. The attention is performed to select important
features across time frames. $\mathbf{\Phi}_{proj}$, performed as the query of the attention, is dot-product with itself, performed as key and value
of the attention, generating attention weights.
The value are weighted averaged using the attention weights, along with residual connection, and
then fed to a GRU encoder. A second self-attention module followed by a linear projection layer is performed to estimate the single-tap speech enhancement filters $\mathbf{W}\in \mathbb{C}^{F\times T \times 3}$ for microphone channel, first stage AEC channel and echo estimation channel, respectively. The loss function in Eq. \ref{loss} is calculated using the combination of SI-SDR\cite{Roux19} in time domain and $l_1$ norm on the spectrum magnitude difference between enhanced signal $\hat{S}_r$ and the target signal $S_r$. The $l_1$ loss on the spectrum magnitude helps to regularize the scale of the output signal.
\begin{equation}\label{loss}
    \mathbf{Loss}_{NeuralEcho} = \mathbf{SISDR}(\hat{s}_r, s_r) + \alpha l_1(\hat{S}_r, S_r)
\end{equation}

\begin{figure}[t]
\vspace{-2cm}
\hspace{-0.3cm}
  \includegraphics[width=170mm, height=130mm]{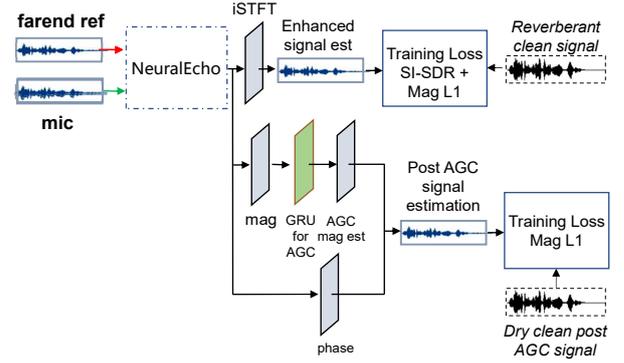}
  \vspace{-6.5cm}
  \caption{The architecture of AGC enabled NeuralEcho model}
  \label{fig:agc}
  \vspace{-0.5cm}
\end{figure}

\begin{table*}[t]
\centering
\caption{\label{table:all} {\it Evaluation on the baseline methods and the proposed methods}}
\vspace{-0.4cm}
\begin{tabular}{l|cccccc}\toprule
 \midrule  \textbf{Method/Metric}  & \textbf{Model Size(M)} & \textbf{SI-SDR}(dB)($\uparrow$) &  \textbf{PESQ}($\uparrow$) &  \textbf{WER}$\%$($\downarrow$) \\
\midrule Reverberent clean reference ($s_r$ in Eq. \ref{signal}) & n/a &$\infty$ & 4.5 &  2.19 \\
Mixture ($y$ in Eq. \ref{signal}) & n/a & -4.23 & 1.79  & 77.12 \\
  \midrule SpeexAEC\cite{Valin16} & n/a & -1.52 &  2.02 & 44.99 \\
   F-T-LSTM \cite{Zhang21} & 8.69  &  10.38 & 2.88  & 15.86 \\
  \midrule NeuralEcho (Section \ref{sec:neuralecho}) & 2.52 & 11.80 & 2.98 & 14.42\\
   Speaker aware NeuralEcho (Section \ref{sec:spk}) & 2.59 &   \textbf{11.98} & 2.99 & 14.06 \\
  NeuralEcho with AGC: pre-AGC(Section \ref{sec:agc}) & 2.52 &  11.62 & \textbf{3.01} & \textbf{13.20} \\
  NeuralEcho with AGC: post-AGC(Section \ref{sec:agc}) & 2.98 &  n/a & n/a & 13.63 \\
\bottomrule
\end{tabular}
\vspace{-0.4cm}
\end{table*}

\vspace{-0.1cm}
\subsection{Speaker Aware Model}\label{sec:spk}
Speech separation has received a lot of attention
in the recent literature using speaker embedding \cite{Delcroix18,Wang19,Ji20, CZhang21}. It improves a certain user’s speech quality not only in multi-talker conditions but also in ambient noises. In this work, we combine the speaker embedding with other acoustic features using feature-wise linear modulation (FiLM) \cite{Perez18} in both stages as shown in Fig. \ref{fig:model}. Similar to \cite{Malley21}, we add a residual connection after FiLM layer in order to ensure that the architecture can perform well when the speaker embedding is absent. Mathematically, this feature fusion module transforms its input acoustic feature $\theta$ and speaker embedding $d_s$ to produce the output feature $\xi$ as follows
\begin{align}
    \hat{\theta} & =  \theta + r(d_s)\odot \theta + h(d_s), \\
    \xi &= \mathbf{LayerNorm}(\hat{\theta}) + \theta,
\end{align}
where $r(\cdot)$ and $h(\cdot)$ are linear projections from 128 dimensional speaker embedding to a 257 dimensional vector to match the size of acoustic feature $\theta$. We trained an ECAPA-TDNN based speaker model \cite{Desplanques20} on 2.8M multi-conditional utterances from 8800 speakers to extract speaker embeddings. The speaker embedding $d_s$, providing a global bias for the target speaker, is provided by averaging speaker vectors over the five enrolled utterances in our setup. 

\subsection{AGC Branch}\label{sec:agc}
In order to adjust and amplify the enhanced speech segment to an intelligible sound level, we add an AGC branch at the output stage of NeuralEcho model, as shown in Fig. \ref{fig:agc}. The enhanced signal magnitude goes through a GRU layer (257 hidden units) to predict the post AGC magnitude, which is then combined with phase of NeuralEcho model's output signal to synthesize the post AGC signal in time-domain. 
There exists a few AGC implementations and guidelines \cite{Archibald08, webrtc}. We use our in-house AGC tool to process the original dry clean data and generate the post-AGC dry clean target signal for the training purpose. Therefore, the output of AGC branch is expected to adjust the sound level as well as suppress reverberation. As pointed in \cite{Luo18}, SI-SDR leads to much slower convergence and worse performance on the dereverberation task, Hence, we use $l_1$ loss on the spectrum magnitude error between the estimated signal and the post-AGC dry clean target signal.

\section{Data and Experimental Setup}\label{exp}
We simulate the single-channel reverberant and noisy dataset using
AISHELL-2 \cite{Du2018} and AEC-Challenge datasets\cite{Sridhar21}. 
We generate
a total number of 10K room impulse responses (RIRs) with random room characteristics and reverberation time (RT60) ranging from 0s to 0.6s using image-source method \cite{Allen79}. Each RIR is a
set consisting of RIRs from near-end speaker, loudspeaker and
background noise locations. We randomly select
RIRs to simulate single-channel AEC dataset. We use clean and
nonlinear distorted versions of far-end signal from AEC-Challenge's synthetic echo set \cite{Sridhar21}. The nonlinear distortions include, but are not limited to:
clipping the maximum amplitude, using a sigmoidal function
\cite{Thiemann13}, and applying learned distortion functions. The signal-to-echo-ratio (SER) is randomly sampled from -10dB to 10dB. In addition, environmental noises are added with signal-to-noise-ratio (SNR) ranging from 0 to 40dB. A total number of 90K, 7.5K, and 2K
utterances are generated for training, validation and testing, respectively.

In this study, we compare our proposed method to SpeexDSP \cite{Valin16}, a signal-processing based
AEC, and F-T-LSTM \cite{Zhang21}, a neural network based single-channel AEC. PESQ \cite{ITU-T01} and SI-SDR \cite{Roux19} serve as the objective evaluation metric for signal quality. The reverberant clean signal $s_r$ is taken as the reference target for computing PESQ and SI-SDR. 
Furthermore, a general-purpose mandarin speech recognition Tencent API \cite{Tencent} is used to test the ASR performance by computing word error rate (WER). For a fair comparison, the echo path delay is not aligned and compensated for the evaluation of all methods. 

\begin{figure}[t]
\vspace{-2.5cm}
\hspace{-0.3cm}
  \includegraphics[width=270mm, height=180mm]{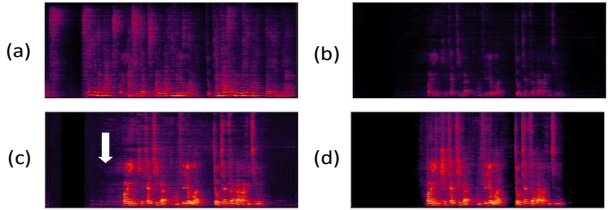}
  \vspace{-13.0cm}
  \caption{(a) Microphone signal, (b) NeuralEcho output, (c) NeuralEcho + AGC post processing, arrow points to the boosted noise (d) NeuralEcho with AGC module: post-AGC output}
  \label{fig:agc_compare}
  \vspace{-0.6cm}
\end{figure}

We compare the proposed methods with the baseline methods in Table \ref{table:all}, where F-T-LSTM is trained by the same dataset. The proposed NeuralEcho method with a much smaller model size outperforms F-T-LSTM in all metrics. We see 13.7\%, 3.5\% and 9.1\% relative improvement in SI-SDR, PESQ and WER, respectively. Due to the loudspeaker non-linear  distortion and the uncompensated echo path delay, the MDF adaptive filter based SpeexAEC does not perform well. To study the significance of speaker information and sound level learning from AGC module, we implemented each of them in Section \ref{sec:spk} and \ref{sec:agc} separately to the NeuralEcho model. With the assist of the target speaker information, the speaker aware NeuralEcho further improves SI-SDR to 11.98dB and PESQ to 2.99, and reduces the WER to 14.06\%. The speaker embedding promisingly helps the model to extract the target speaker from its mixture with far-end speakers and other near-end noises. Equipped with the AGC module, the NeuralEcho model is able to predict the reveberant clean speech (pre-AGC) as well as the post-AGC dry clean speech (post-AGC) through a multi-task learning. During the inference stage, the pre-AGC branch only depends on the original NeuralEcho model part. The results on the pre-AGC output demonstrates the benefits of adding AGC branch in the NeuralEcho model and the multi-task training of the model. Despite a little regression on SI-SDR compared to the best value achieved by speaker aware NeuralEcho model, it achieves the best PESQ and WER. Particularly, WER is reduced by 16.8\% compared to F-T-LSTM method and reduced by 8.5\% compared to the NeuralEcho model in Section \ref{sec:neuralecho}. The SI-SDR and PESQ are not computed for the post-AGC output as its training target (AGC processed dry clean signal) is not aligned with the reverberant clean reference signal which is used for computing the two metrics. The pre-AGC output performs better than post-AGC output on WER, possibly due to its less residual noise amplification from AGC and less speech distortion from dereverberation, which is favored by the speech recognition API. Nevertheless, the WER on post-AGC output achieves 13.63\%, which is the second best result.

We proved the importance of the joint training with AGC module in Table \ref{table:agc}. Directly training NeuralEcho model with AGC processed clean data leads to a much worse WER (16.53\%). Such training target causes slow convergence. Processing the NeuralEcho output by the in-house AGC tool is not able to improve the WER. The defect in the output of NeuralEcho model can likely be amplified by the downstream AGC module as shown in Fig. \ref{fig:agc_compare}. The best method, NeuralEcho with AGC branch, demonstrates the benefit of joint training AGC and NeuralEcho as a unified model, which achieves the best WER and amplifies the sound level like what an typical AGC algorithm performs ((d) in Fig. \ref{fig:agc_compare}).

\begin{table}[t]
\centering
\caption{\label{table:agc} {\it The importance of joint training with AGC}}
\vspace{-0.2cm}
\begin{tabular}{l|ccc}\toprule
 \midrule  \textbf{Method/Metric}   &   \textbf{WER}$\%$($\downarrow$) \\
  \midrule NeuralEcho & 14.42\\
   NeuralEcho trained with post AGC clean data & 16.53 \\
   NeuralEcho + AGC post processing  &    14.75 \\
  NeuralEcho with AGC: post-AGC   &  \textbf{13.63} \\
\bottomrule
\end{tabular}
\vspace{-0.6cm}
\end{table}

\section{Conclusions}\label{con}
In this paper we designed a unified echo suppression and speech enhancement model, namely NeuralEcho. Compared to the existing deep learning based approaches, our contribution is three-fold, i) leveraging second order statistics across signal channels, and time and frequency axes leads to significant AEC feature learning, ii) we propose a self-attentive RNN neural network to perform AEC on microphone channel, intermediate AEC processed channel and echo estimation channel, iii) the proposed model is further developed to work with speaker information and perform AGC for better target speaker enhancement. Experimental results showed that the proposed NeuralEcho and its derivatives yield a significant improvement to the conventional signal processing based method and the state-of-the-art neural network method in terms of both objective audio quality and speech recognition accuracy. 

\newpage
\bibliographystyle{IEEEtran}
\bibliography{mybib}

\end{document}